\documentclass[twocolumn]{aa}
\usepackage[varg]{txfonts}

\usepackage{graphicx}
\usepackage{natbib}
\usepackage{color}
\usepackage{subfig}
\usepackage[switch]{lineno}
\usepackage{xspace}
\usepackage{microtype}
\usepackage{amssymb}
\usepackage{mathtools}
\usepackage{multirow}
\def \link_col{blue}
\usepackage{xspace}
\usepackage{url}
\usepackage{wasysym}
\usepackage{caption}

\newcommand{\be}{\begin{equation}}
\newcommand{\ee}{\end{equation}}

\newcommand{\voy}{{\sl Voyager 1}\xspace}

\newcommand{\gray}{{\rm $\gamma$-ray}\xspace}
\newcommand{\grays}{{\rm $\gamma$-rays}\xspace}
\newcommand{\alp}{{\rm $\alpha$}}

\newcommand{\mn}{{${\rm MeV/nuc}$}\xspace}
\newcommand{\gn}{{${\rm GeV/nuc}$}\xspace} 

\def \ec{\element[][12]{C}\xspace}
\def \en{\element[][14]{N}\xspace}
\def \eo{\element[][16]{O}\xspace}
\def \ene{\element[][20]{Ne}\xspace}

\def \esi{\element[][28]{Si}\xspace}

\def \efe{\element[][56]{Fe}\xspace}

\begin{document}

\title{Nuclear de-excitation lines as a probe of low-energy cosmic rays}
\author{Bing Liu\inst{1,2,3,4}
\and Rui-zhi Yang\inst{1,2,3 \dagger}
\and Felix Aharonian\inst{5,6,7}
}
\institute{Department of Astronomy, School of Physical Sciences, University of Science and Technology of China, Hefei, Anhui 230026, China  $^\dagger$ yangrz@ustc.edu.cn
\and  CAS Key Laboratory for Research in Galaxies and Cosmology, University of Science and Technology of China, Hefei, Anhui 230026, China 
\and School of Astronomy and Space Science, University of Science and Technology of China, Hefei, Anhui 230026, China 
\and Key Laboratory of Modern Astronomy and Astrophysics (Nanjing University), Ministry of Education, Nanjing 210093, China
\and Dublin Institute for Advanced Studies, 31 Fitzwilliam Place, Dublin 2, Ireland 
\and Max-Planck-Institut f\"ur Kernphysik, P.O. Box 103980, D 69029 Heidelberg, Germany 
\and National Research Nuclear University MEPhI, Kashirskoje Shosse, 31, 115409 Moscow, Russia
}

\abstract{
Low-energy cosmic rays (LECRs) contribute substantially to the energy balance of the interstellar medium. They play also significant role in the heating and chemistry of gas, and, consequently, on the star formation process. Because of the slow propagation coupled with enhanced energy losses of subrelativistic particles, LECRs are concentrated around their acceleration sites. LECRs effectively interact with the ambient gas through nuclear reactions. Although these processes are energetically less effective compared to heating and ionization,  they are extremely important from the point of view of\ nuclear de-excitation lines, which carry unique information about LECRs. We present results on production of de-excitation lines 
combining the numerical treatment of nuclear reactions using the code TALYS, with the 
propagation and energy losses of LECRs. }

\maketitle

\section{Introduction}
\label{sec:intro}

The energy density of Galactic cosmic rays (CRs), $w_{\rm CR} \approx 1 ~\rm eV/cm^3$, is comparable to the energy density contributed by the interstellar magnetic fields and thermal gas. CRs play an essential role in the process of star formation through the heating and ionization, initiating several crucial chemical reactions in the dense cores of molecular clouds \citep{papadopoulos}. 

The flux and spectra of CRs inside the Solar System have been  measured with unprecedented accuracy by space detectors such as  PAMELA\footnote{A Payload for Antimatter Matter Exploration and Light-nuclei Astrophysics} \citep{pamela2013,pamela2014}, AMS-02\footnote{The Alpha Magnetic Spectrometer} \citep{ams2015p,ams2015he}, CREAM\footnote{The Cosmic Ray Energetics and Mass experiment} \citep{cream2017}, DAMPE\footnote{The Dark Matter Particle Explorer} \citep{dampe2019SciA}, and CALET\footnote{The CALorimetric Electron Telescope} \citep{calet2020pan}. Most of these direct measurements from Earth focus on the CR spectra above a few GeV/nuc, below which the fluxes are strongly affected by solar modulation.
A few years ago, the {\sl Voyager 1} satellite passed through the termination shock and measured the low-energy CR (LECR) spectra (from several MeV/nuc up to hundreds of MeV/nuc) beyond the Solar System \citep{Cummings2016}. The measurements outside the heliosphere are thought to be free of solar modulation and thus may represent the LECR spectra in the local interstellar medium. However, this information cannot be extrapolated to other parts of the Galaxy,  in which the CRs are not homogeneously distributed \citep{Aharonian2018,Johannesson2018,Baghmanyan2020}. Furthermore, the flux of LECRs can vary dramatically on a smaller scale because of higher energy losses and propagation effects,  and the flux is usually much higher around possible CR acceleration sites. This is indirectly supported by studies of the interstellar ionization rates (in particular, see, e.g., \citealp{Indriolo2009,Indriolo2012,Indriolo2015}). These researches also argued for an LECR component in addition to the standard contribution by supernova remnants (SNRs), which is also supported by the observation of primary Be \citep{Tatischeff2011}.

In this regard, \grays produced in interactions of CRs with the ambient gas can be used as a unique tool to study the spectral and spatial distributions of CRs throughout the Milky Way. For CRs whose kinetic energy exceeds the kinematic threshold of $\pi$-meson production, $E_{\rm th}\simeq 280$\,MeV/nuc,  the best $\gamma$-ray energy band for exploration is 0.1 - 100\,GeV because of the copious production of $\pi^0$-decay $\gamma$-rays and the potential of the currently most sensitive detector, {\sl Fermi} LAT \citep{Atwood2009fermilat}. At energies below this kinematic threshold,  the nuclear de-excitation lines provide the most straightforward information about the LECR  protons and nuclei \citep[e.g.,][]{Ramaty1979,Murphy2009}. 

In this paper, we treat the production of nuclear de-excitation lines by combining the recent advances in the modeling of nuclear reactions with the propagation and energy losses of subrelativistic and transrelativistic CRs. In Sec.~\ref{sec:pspec} we calculate the distributions of LECRs near the sources, taking the processes of diffusion and energy losses into account. In Sec.~\ref{sec:nlines} we describe the method for calculating the emissivity of de-excitation \gray lines using the code TALYS  \citep{talys2008}. We then  present the results and discuss their implications. Finally, we summarize the main results in Sec.~\ref{sec:sum}.

\section{Propagation of LECR protons}
\label{sec:pspec}
\begin{figure}
\centering
\includegraphics[width=0.45\textwidth]{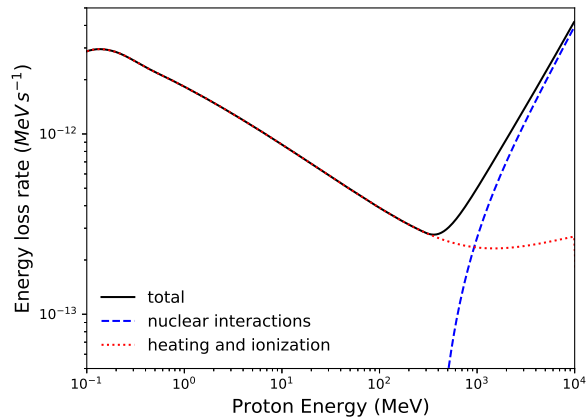}
\caption{Energy loss-rate as a function of kinetic energy for protons colliding with atomic hydrogen. The number density of atomic hydrogen $n$  is assumed to be $1\,{\rm cm}^{-3}$. The dashed blue line indicates the loss rate due to p-p inelastic interactions, and the dotted red line represents the energy losses caused by heating and ionization. }
\label{fig-te}
\end{figure}

The energy losses of CR protons with kinetic energies above $\sim$1\,GeV are dominated by inelastic nuclear reactions (see Fig.\ref{fig-te}), and the corresponding cooling  weakly depends on energy. As a result, the propagation has very little effect on the initially injected CR spectral shape. At lower energies, the losses are dominated by the ionization and heating of the ambient medium. In this regime, as shown in Fig.\ref{fig-te},  the cooling time is energy dependent, thus the propagation can significantly distort the initial spectrum. On the other hand, because of the energy-dependent particle diffusion, LECRs propagate more slowly than relativistic CRs.  
Consequently, the spatial distribution of LECRs should be much more inhomogeneous than that of high-energy CRs. LECRs are expected to be concentrated in the vicinity of their accelerators with energy distributions that strongly depend on the distance to their production sites.  The propagation of LECRs in the proximity of their sources should therefore be treated with great care and depth.

As introduced in \citet{Padovani2009}, $L=-dE/dN_{\rm H_2}$, represents the energy loss-rate per column density  $dN_{\rm H_2}$, then the ionization cooling time for protons colliding with a medium atomic hydrogen density $n$ can be estimated by $\tau\sim\frac{E}{L\,n\,\beta\,c}$, where $\beta =v/c$. 
For protons with kinetic energy 10\,MeV that collide with ${\rm H_2}$, $L\simeq 2\times10^{-16}\,\rm eV\,cm^2$  (see Fig.7 in \citealt{Padovani2009}). The cooling time of 10\,MeV protons in a dense environment with $n\geq 100\,\rm cm^{-3}$ therefore is  about several thousand years, much shorter than the duration of typical CR accelerators such as SNRs.  Therefore we adopt the steady-state solution for continuous injection to estimate the spectra of LECRs. To be more specific, we assume a continuous injection of protons from a stationary point source. Then the steady-state energy and radial distribution distribution of the LECR protons is obtained by application of the analytical solution derived by \citet{Atoyan1995}, 
\begin{equation}
\label{equ1}
F_{\rm p}(r,E)=\frac{1}{8\pi^{3/2}P(E)}\int^{\infty}_{E} \frac{Q(x)}{[\Delta u(E, x)]^{3/2}}
\times {\rm exp}\left(-\frac{r^2}{4\Delta u(E,x)}\right){\rm d}x
.\end{equation}
Here $Q(E)$ represents the CR injection rate, $P(E)$ is the energy loss-rate of CR protons summarized in \citet{Padovani2009}, $r$ represents the radial distance to the source, and $\Delta u(E,x) = \int^{x}_{E} \frac{D(E')}{P(E')}dE'$, where $D(E)$ is the energy-dependent diffusion coefficient of CRs. In general, this is a quite uncertain parameter that depends on the level of turbulence in the environment. In the Galaxy, it is derived from observations of secondary CRs. We adopted the diffusion coefficient derived for Galactic CRs at energies above 1 GeV \citep{Strong2007review} and assumed that it can be extrapolated to low energies down to 1\,MeV: $D(E)=4\times10^{28} \chi (E/1 \rm \ GeV)^{0.5}\ {\rm cm^2\,s^{-1}}$. 
Because the turbulence level near the CR sources is high, the diffusion could be much slower \citep{Malkov2011,DeAngelo2018}.  Effects like this have previously been observed in the \gray band. The electron halos near pulsars reveal a small diffusion coefficient \citep{hawc_geminga, fermi_geminga}.  To take this effect into account, we considered a broad range of the parameter $\chi=1,0.1,0.01$. Here the total proton injection rate $Q$, integrated from 1\,MeV to 100\,MeV,  and the distance of the source $d$, are fixed by the value of parameter $Q/d^2=10^{38}\,\rm erg\,s^{-1}\,kpc^{-2}$.

The injection spectrum of CRs depends on the acceleration mechanism and  the conditions inside the accelerator such as the turbulence level and the magnetic field. Typically (although not always), in the case of diffusive shock acceleration, for instance, the distribution of accelerated particles can be presented as a power law in momentum $p$,  $Q(p)=Q_0p^{-s}$ \citep[see, e.g., ][]{amato14}.
In this paper, it is more convenient to write the distribution in terms of kinetic energy $E$, $Q(E)=Q_0p^{-s}/\beta$, where  $Q_0$ is the normalization parameter derived from the power of the CR source. 
By solving Eq.\ref{equ1}, we obtain the steady-state distribution of protons at different radial distances $r$ to the source.

\begin{figure*}
\centering
\includegraphics[width=0.45\textwidth]{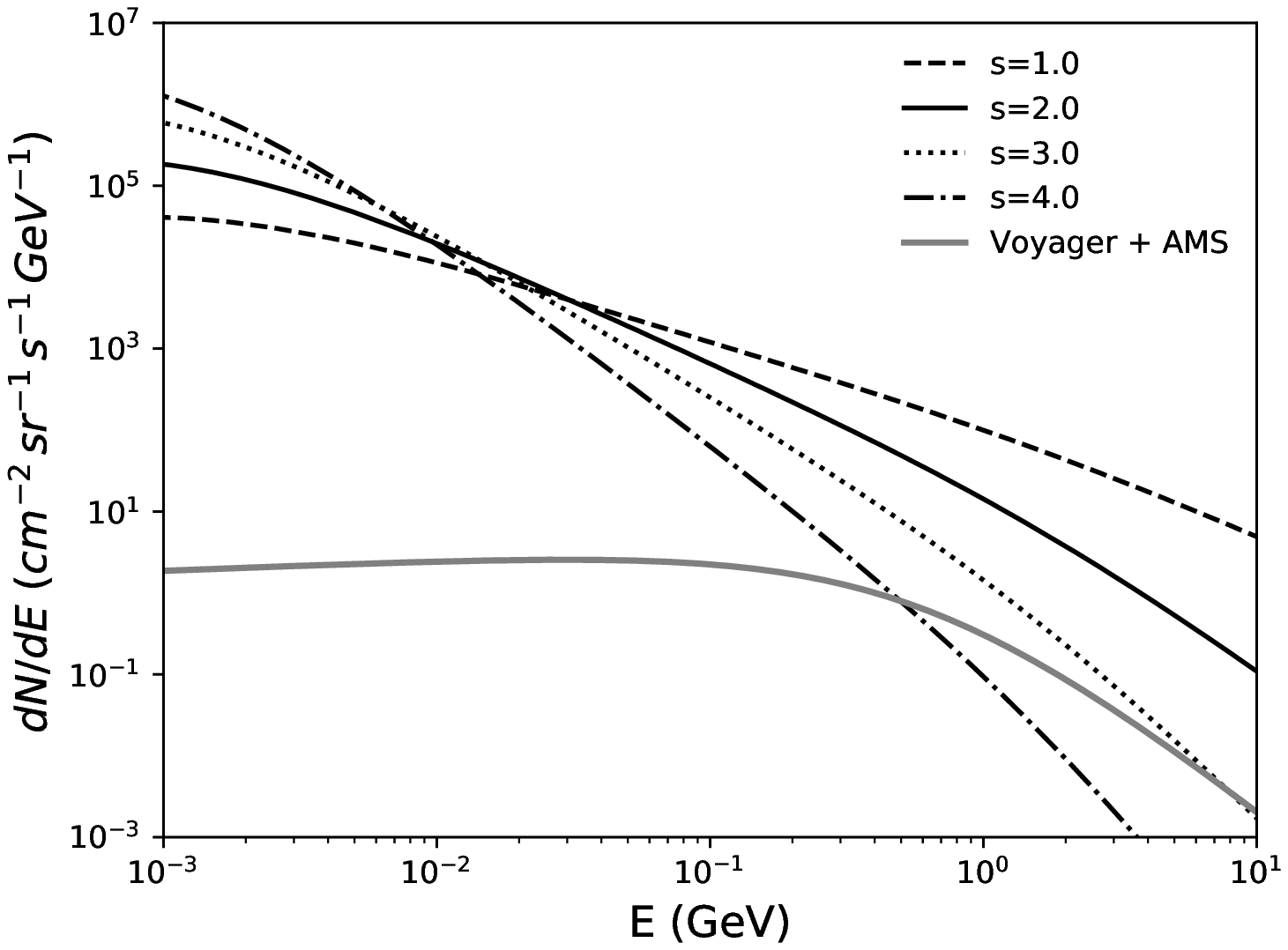}
\includegraphics[width=0.45\textwidth]{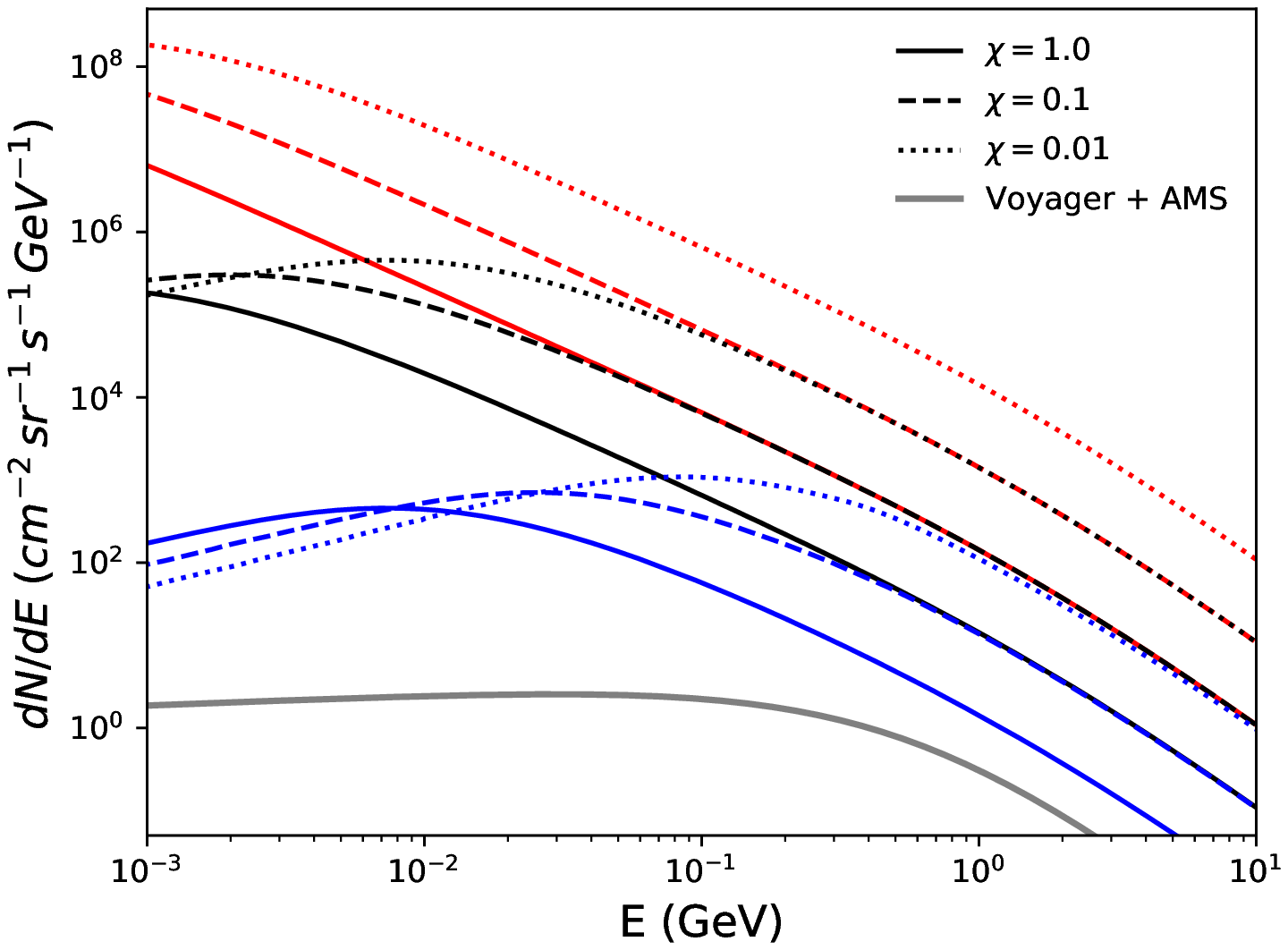}
\caption{ Calculated proton spectra under different assumptions. Left panel: Proton spectra at the radical distance $r=10$\,pc from the hypothetical CR source with various injection spectral indices, in which $s=$1.0 (dashed line), 2.0 (solid line), 3.0 (dotted line), and 4.0 (dash-dotted line), assuming a diffusion coefficient parameter $\chi=1.0$. 
Right panel: Proton spectra with injection spectral index $s=2.0$ under various assumptions of diffusion coefficients (solid lines for $\chi=1.0$, dashed lines for $\chi=0.1$, and dotted lines for $\chi=0.01$) at different radial distances, where $r=$ 1\,pc (red), 10\,pc (black), or 100\,pc (blue).  
In both panels, $Q/d^2=1\times10^{38}\,\rm{erg\,s^{-1}\,kpc^{-2}}$ and $n=1\,\rm cm^{-3}$ are assumed, and the gray lines represents the fit of the CR proton intensity from \citet{Phan2018} based on the measured fluxes of local CRs reported by {\sl Voyager 1} \citep{Cummings2016} and AMS-02 \citep{ams2015p}.}
\label{fig-pspec}
\end{figure*}

The energy spectra of CR protons  diffusing from the site of their acceleration are shown in Fig.\ref{fig-pspec}. In the left panel of Fig.\ref{fig-pspec}, the curves are calculated for four 
injection spectra of protons with power-law index $s=$1.0, 2.0, 3.0, and 4.0, in which the radial distance $r=10\,\rm pc$, the medium density $n=1\,\rm cm^{-3}$ , and the parameter $\chi=1$. 
In the right panel of Fig.\ref{fig-pspec}, we present the CR spectra for the injection spectrum with $s=2.0$ at various distances, in which $r=$1\,pc, 10\,pc, and 100\,pc. By changing the diffusion coefficient, that is, the  parameter $\chi=$1, 0.1, and 0.01, we find that for a small diffusion coefficient, the energy spectra of protons become very hard, especially at large distances from the source.  

\begin{figure*}
\centering 
\includegraphics[width=0.45\textwidth]{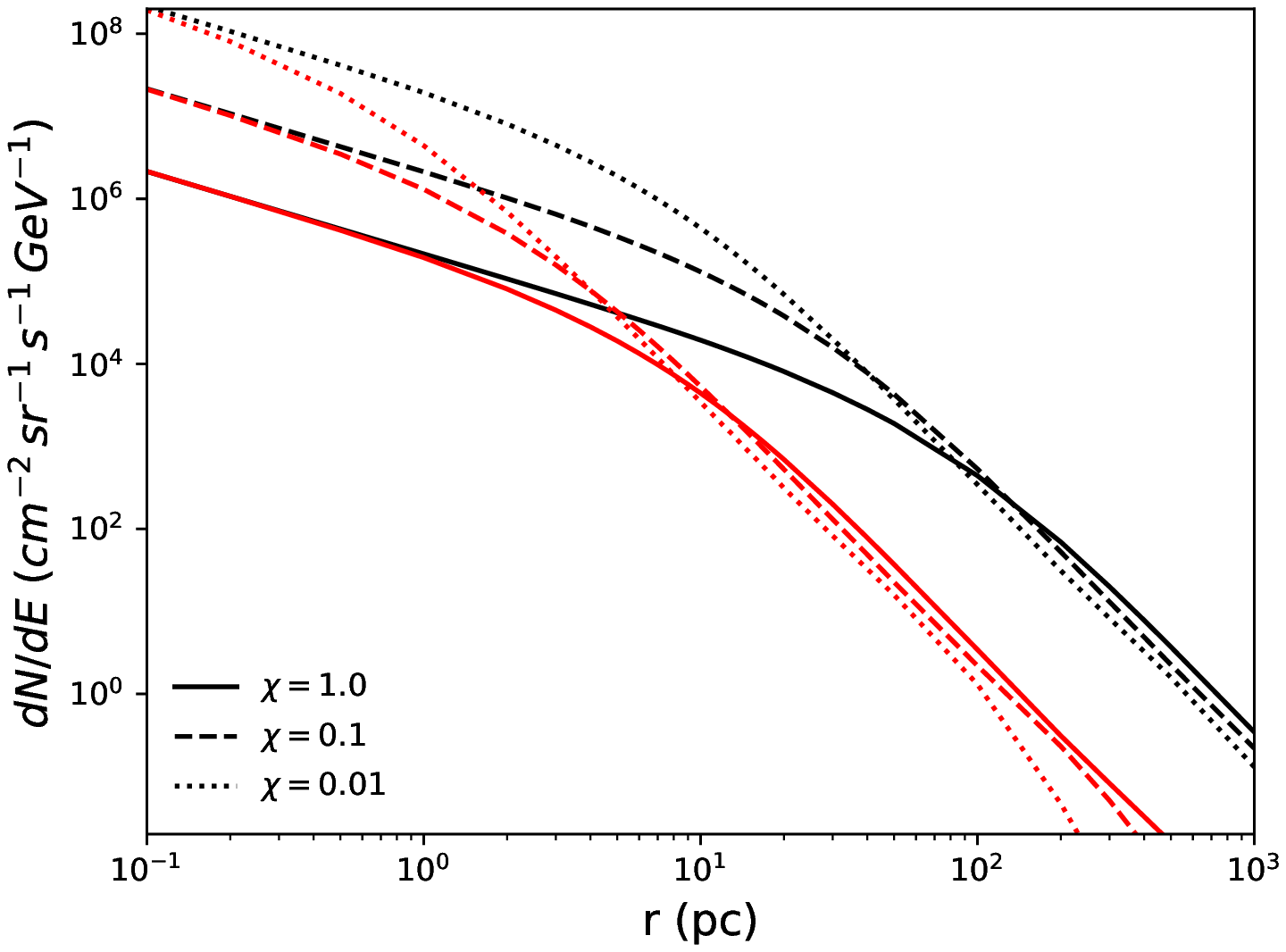}
\includegraphics[width=0.45\textwidth]{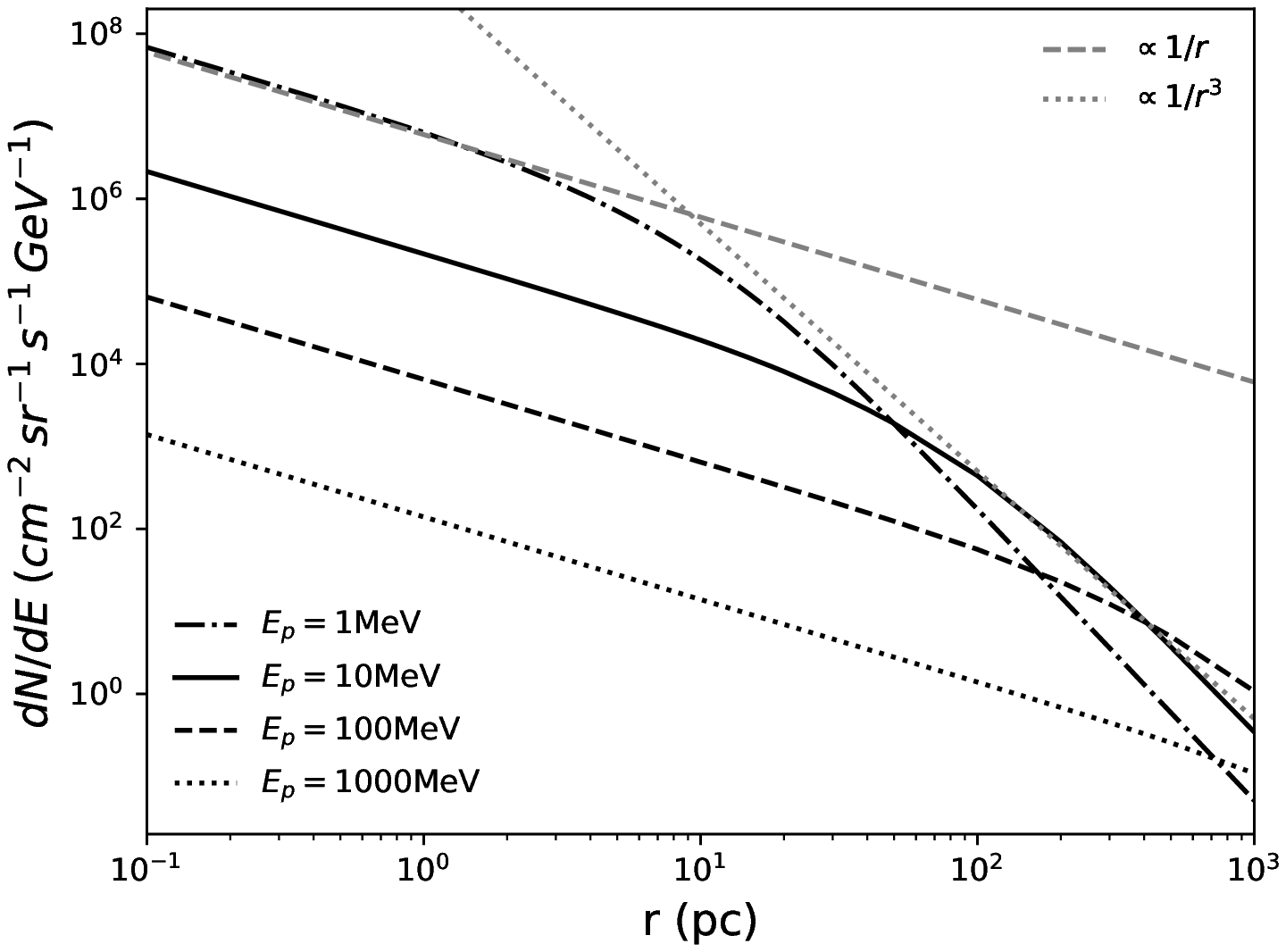}
\caption{  Radial distribution of protons with certain kinetic energies.  
 Left panel: Flux of 10\,MeV protons as a function of radial distance $r$ under different assumptions of medium density (black lines for $n=1\,\rm cm^{-3}$, and red lines for $n=100\,\rm cm^{-3}$) and diffusion coefficient (solid lines for $\chi=1.0$, dashed lines for $\chi=0.1$, dotted lines for $\chi=0.01$).
Right panel: Flux of protons with different kinetic energies $E_p$ as a function of radial distance $r$, assuming $n=1\,\rm cm^{-3}$, and $\chi=1$. For comparison, two radial profiles $1/r$ (dashed gray line) and $1/r^3$ (dotted gray line) are also shown. In both panels, we assume $s=2.0$ and $Q/d^2=1\times10^{38}\,\rm{erg\,s^{-1}\,kpc^{-2}}$.
}
\label{fig-pspec1}
\end{figure*}

In Fig.~\ref{fig-pspec1} we show the radial dependence of proton fluxes. In the left panel, we show the effect of the  diffusion coefficient on the flux of 10\,MeV protons. At small distances from the source, their proton flux is significantly higher for slow diffusion, but at large distances, the flux drops as the cooling becomes an important factor. The contrast of the spectra under different assumptions of $n$ (the red and black lines in the left panel of Fig.\ref{fig-pspec1}) also shows the effect of the ambient  gas density on the CR spectrum.  As expected, because of enhanced energy losses, the radial distribution of the CR fluxes become sharper than $1/r$ at larger distances, $r \geq 10$\,pc. In the right panel, we show the $r$-profiles for different CR energies calculated for the parameter $\chi=1$.  
For the radial profile of the CR, the flux at 1\,GeV is close to $1/r$. This is consistent with predictions for the continuous injection in \citet{Atoyan1995}, as long as the energy losses are negligible. At low energies, $E\le 1\,\rm GeV$ because of the high energy losses and slow propagation, we see a stronger radial dependence of the CR flux. Namely, at small distances, the flux follows the  $1/r$ profile, but at larger distances the cooling starts to play significant role and causes the radial distribution to become as steep  as $1/r^3$. For 1\,MeV protons, the transition between $1/r$ and $1/r^3$ takes place at just several pc.  Last but not least, we note that the hardening of the spectrum of CR protons at low energies in the local interstellar medium, as shown by the gray line in Fig.~\ref{fig-pspec}, fitted by \citet{Phan2018} using data from {\sl Voyager 1} and AMS-02 detectors), can be naturally explained by the propagation and energy losses.

\begin{figure}
\centering
\includegraphics[width=0.45\textwidth]{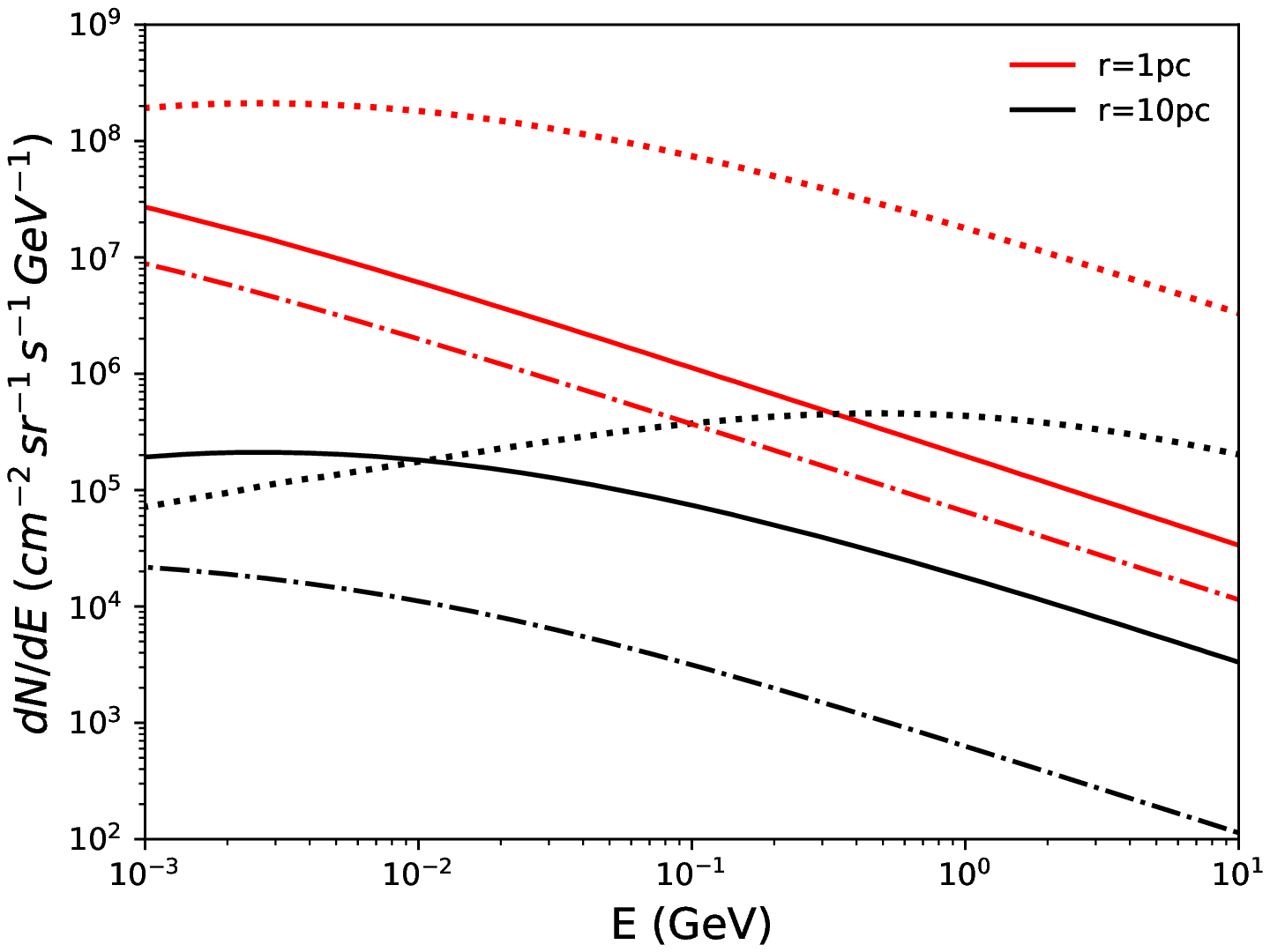}
\caption{Proton spectra at different radial distances, in which $r=1$\,pc (red) or 10\,pc (black). Assuming $s=2.0$, $n=1\,\rm cm^{-3}$, and $Q/d^2=1\times10^{38}\,\rm{erg\,s^{-1}\,kpc^{-2}}$, the solid ($\chi=1.0$) and dotted lines ($\chi=0.01$) are calculated for the diffusion 
using Eq.(\ref{equ1}), and the dash-dotted lines are calculated for advection using  Eq.(\ref{equ2}) for $V=1000\,\rm km/s$.} 
\label{fig_cov}
\end{figure}

In the derivation of Eq.(1), only the  diffusion and energy loss of protons are taken into account. Meanwhile, the  advection may also play a non-negligible role in the vicinity of CR sources. In this case, an advection term $V \frac{\partial F_{\rm p}(r,E)}{\partial r}$ should be added to the propagation equation, where $V$ is the advection velocity. By dimensional analysis, the advection dominates diffusion in the region $r > D/V \sim 30\,{\rm pc} \frac{D}{10^{27}~\rm cm^2 s^{-1}}/\frac{V}{100~ \rm km/s}$. The Alfv$\acute{e}$nic velocity in the interstellar medium  is estimated as several $\rm km/s$ assuming the density of $\sim1\,\rm cm^{-3}$ and magnetic fields of about $3\,\rm \mu G$  \citep[see, e.g., ][]{han17}. In this case the advection can be neglected close to the source. However, for a strong outflow of gas or a stellar wind near the source, with a speed as as high as $1000~\rm km/s$  \citep[see, e.g.,][]{santamar06}, at distances $r \geq 0.1$\,pc, the advection can dominate the diffusion, resulting in the steady-state solution 
\begin{equation}
\label{equ2}
F_{\rm p}(r,E)=\frac{Q(E_0)}{4\pi r^2 \beta c}, 
\end{equation}
where $E_0$ is obtained from the equation $\frac{r}{V}=\int_{E}^{E_0}\frac{dE'}{P(E')}$. The corresponding CR fluxes are shown in Fig.\ref{fig_cov}. The fast advection results in the strong  suppression of CR flux even compared with the fast diffusion case ($\chi=1$) in the vicinity of the source. 

Moreover, in the calculation above we assumed that the power-law diffusion coefficient extends to low energy. Recent analyses show indications of a low-energy break in the diffusion coefficient  in the ISM \citep{Vittino2019,Weinrich2020}, which was also predicted in \citet{Ptuskin2006} because the damping of CRs terminates the cascade of turbulence and induces faster diffusion for LECRs. However, it is not straightforward to adopt this scenario in the vicinity of the CR sources, where the external turbulence is also stronger by an order of magnitude. The detailed calculation requires a self-consistent treatment of the CR propagation and magnetic turbulence development and damping near the CR sources, which need further investigations.

\section{Rates and spectral features of the de-excitation $\gamma$-ray line emission}
\label{sec:nlines}

\subsection{Method for estimating the $\gamma$-ray line emission  }
\label{subsec:method}
The interactions of LECRs with the surrounding gas excite nuclei that belong to LECRs and to the ambient medium. The almost prompt de-excitation of these nuclei leads to the MeV
$\gamma$-ray line emission. The main production of $\gamma$-ray lines proceeds 
through (1) energetic protons and $\alpha$-particles as projectiles interacting with  heavier elements of the ambient gas, that is, {\sl \textup{direct}} reactions, and (2) interactions 
of heavy nuclei of LECRs with the hydrogen and helium of the ambient gas, that is, {\sl \textup{inverse}} reactions.  Both channels were taken into account in the calculations of the \gray  emission.  For simplicity, we only considered the most abundant stable isotope of elements, including C, N, O, Ne, Na, Mg, Al, Si, S, Ca, Ar, Fe, and Ni, 
and disregarded the isotopes with lower  abundance, such as \element[][3]{He}, \element[][13]{C,} and \element[][22]{Ne}. The abundance of these element species in the LECRs were extracted from Table~3 of \citet{Cummings2016}. For the composition of the ambient medium, we used the recommended present-day solar abundances of \citet{Lodders2010}. These data are compiled in Tab.\ref{tab-ab}.
Moreover, we assumed that the injected energy spectra of energetic \alp\ and  other species have the same shape as that of CR protons, $F_{i}(r,E)\propto F_{\rm p}(r,E)$, where $E$ is expressed as kinetic energy  per nucleon, 
and we calculated their propagated spectra using the same method as described in Sec.\ref{sec:pspec}. The energy loss-rate of heavy nuclei can be significantly different from that of protons. To account for these differences, the energy loss-rate for each nucleus species including ionization and nuclear interaction were derived from formulae in \citet{Mannheim1994} and cross sections in  \citet{Sihver1993}.

To calculate  emissivities of the de-excitation $\gamma$-ray line lines,  we used the code TALYS  (version 1.95) \footnote{\url{https://tendl.web.psi.ch/tendl_2019/talys.html}}, which is a flexible and user-friendly program aimed at a complete and accurate simulation of nuclear  reactions \citep{talys2008}. The newly updated version of TALYS provides adequate precision of cross sections at projectile energies $<$1 GeV \citep{Koning2014}. For a better match with the experiment data, we modified the deformation files of \en, \ene, and \esi  using the results of \citet{Benhabiles2011}. Then, following the approach of \citet{Murphy2009}, we divided the $\gamma$-ray emission into three categories: the explicit lines, the quasi-continuum consisting of discrete lines,  and the continuum.  All the discrete lines with the TALYS-calculated cross sections with $>$10 mbarn were selected as explicit lines. Those with smaller cross sections were treated as quasi-continuum.
The continuum component produced by the so-called direct, pre-equilibrium, and compound reactions (see the TALYS user manual for a detailed description) was considered as well. Because in the updated version of TALYS the projectile energy relevant to  reactions with involvement of \alp-particles  is limited to 250 \mn,  we extrapolated these line production cross sections to 1 \gn and  assumed that the continuum production cross section remains constant when the projectile energies exceed 250 \mn.  The production cross sections of the specific lines listed in the compilation of \citet{Murphy2009} were extracted from Table 3 therein. For each explicit line, we derived the line profile according to the method proposed in \citet{Ramaty1979}.

\subsection{Results and discussion}
\label{subsec:dis}
\begin{figure}
    \centering
    \includegraphics[width=0.45\textwidth]{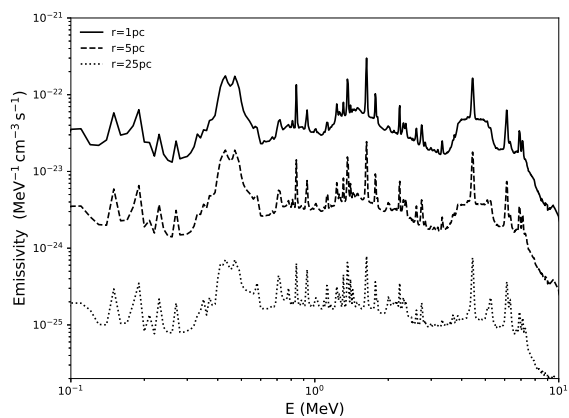}
   \caption{ Calculated \gray emissivity at different radial distances from the hypothetical CR source, in which $r=1$\,pc (solid line), 5\,pc  (dashed line), and 25\,pc (dotted line),
   assuming $Q/d^2=1\times10^{38}\,\rm{erg\,s^{-1}\,kpc^{-2}}$, $d=1$\,kpc, $s=2.0$, and $n=100\,\rm cm^{-3}$.}
    \label{fig5-l}
\end{figure}

\begin{figure}
    \centering
    \includegraphics[width=0.45\textwidth]{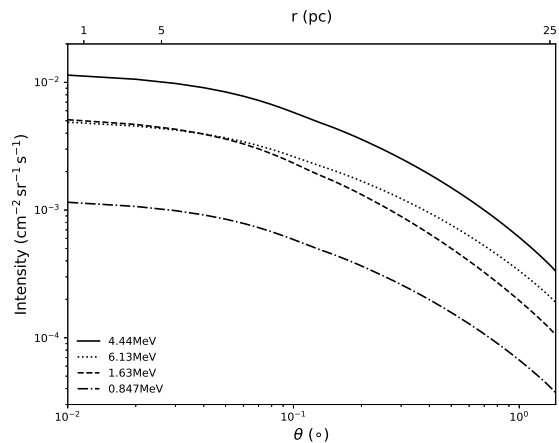}
   \caption{  Integrated flux of strong narrow lines at 4.44 MeV (solid line), 6.13 MeV (dotted line), 1.63 MeV (dashed line), and 0.847 MeV (dash-dotted line), as a function of angular distance $\theta$ (or radial distance $r$) from the hypothetical CR source.  $Q/d^2=1\times10^{38}\,\rm{erg\,s^{-1}\,kpc^{-2}}$, $d=1$\,kpc, $s=2.0$, and $n=100\,\rm cm^{-3}$ are assumed for the calculation.}
    \label{fig5-r}
\end{figure}

\begin{figure*}
\centering
\includegraphics[width=0.45\textwidth]{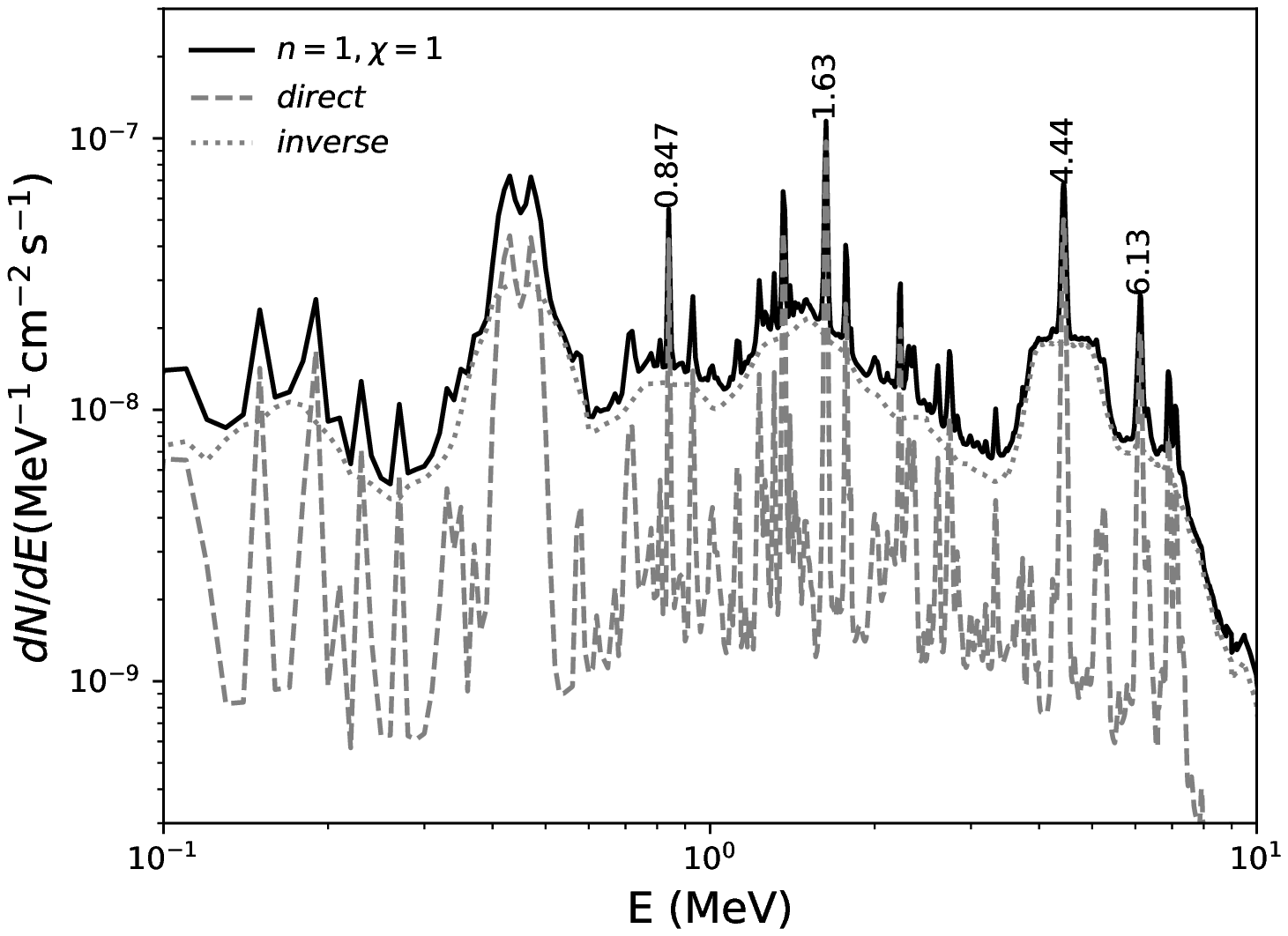}
\includegraphics[width=0.45\textwidth]{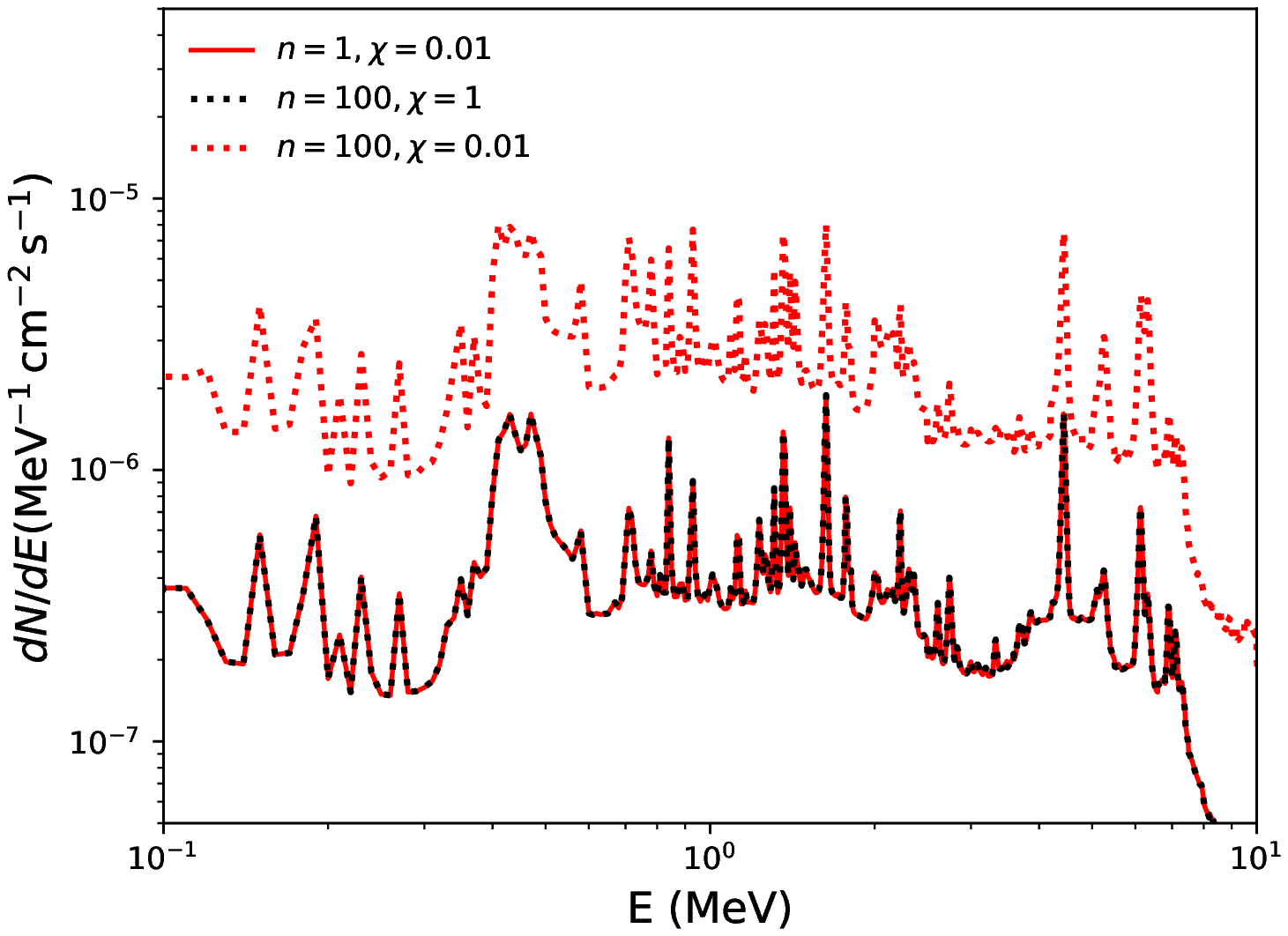}
\caption{ Comparison of the integrated \gray differential fluxes for different $n$ ($1\,\rm cm^{-3}$ or $100\,\rm cm^{-3}$) and $\chi$ (1.0 or 0.01).  In the left panel, the solid line show the total $\gamma$-ray flux integrated  within $r=25$\,pc from the hypothetical CR source, assuming $\chi=1.0$ and $n=1\,\rm cm^{-3}$, 
and the dashed and dotted lines correspond to the contributions from the direct and inverse processes, respectively. In the right panel, the lines  represent the total integrated \gray line emission under different assumptions of $n$ and $\chi$. In both panels, we assume $Q/d^2=1\times10^{38}\,\rm{erg\,s^{-1}\,kpc^{-2}}$, $s=2.0$.}
    \label{fig6}
\end{figure*}

For the given compositions of LECRs and the ambient gas,   the detectability of the  $\gamma$-ray lines, that is, their flux integrated over the entire region occupied by LECRs, depends on several parameters. First of all, the LECR injection rate of source $Q$ versus its distance $d$, which could be represented by the parameter $Q/d^2$, the density of the ambient gas $n$,  the speed of propagation of LECRs, and  their original spectral shape. With the method described above, we first calculated the emissivity of the \gray line emission at different radial distances $r$ to a hypothetical CR source that continuously injects CRs with a power-law index $s=2$ and $Q/d^2=10^{38}\,\rm{erg\,s^{-1}\,kpc^{-2}}$ into the surrounding medium of $n=100\,\rm cm^{-3}$. The results, as shown in  Fig.\ref{fig5-l}, show a significant effect of the propagation process on the emissivities of the \gray line emission.   
We then integrated the intensity of several representative narrow lines along the line of sight at different angular distances $\theta$, assuming the distance of the source $d$ is 1\,kpc. As shown in  Fig.\ref{fig5-r}, the fluxes drop sharply with $\theta$ and become almost negligible at $\theta\sim{1.5}^\circ$ , which corresponds to a radial distance $r\sim 25$\,pc.  Unfortunately, the angular resolution of MeV $\gamma$-ray detectors, including those planned for the foreseeable future, such as e-Astrogam \citep{e-astrogam2018} and AMEGO \citep{amego2019}, is quite limited, about ${1.5}^\circ$ , thus the angular distribution of these lines can be detected only for very nearby objects, $d\ll1$\,kpc.  In the following discussion we therefore treat the \gray emission around the injection point as a point-like source, and calculate the differential flux of this \gray source by integrating the \gray emission within 25\,pc from the injection site,
\begin{equation}
\label{equ3}
F_{\rm \gamma}(E_{\rm \gamma})=\frac{1}{4\pi d^2}\int^{r_{25}}_{0} 4\pi r^2\,I(E_{\rm \gamma},r){\rm d}r 
,\end{equation}
where $I(E_{\rm \gamma},r)$ is the \gray emissivity at the distance $r$.
In the left panel of Fig.~\ref{fig6}, we show the differential fluxes of $\gamma$-rays calculated for $s=2.0$, $Q/d^2=10^{38}\,\rm{erg\,s^{-1}\,kpc^{-2}}$, $n=1\,\rm cm^{-3}$, and $\chi=1$.  The brightest lines are 0.847, 1.634, 4.44, and 6.13 MeV lines resulting from the de-excitation of \efe, \ene, \ec, and \eo, respectively, from their first or second excited state to ground state. The  {\it \textup{direct}} (narrow) and {\it \textup{inverse}}  (broad)  components linked to the nuclei of LECRS and the ambient gas both contribute substantially to the total flux. The {\it \textup{direct}} processes contribute to the narrow line structures, while the {\it \textup{inverse}} processes are responsible for the broad lines and thus contribute mostly to the continuum emissions. In the right panel of Fig.~\ref{fig6}, we compare the fluxes of $\gamma$-ray emission integrated within 25\,pc radial distance from the source for two different gas densities $n=1\ \rm cm^{-3}$ and $n=100\ \rm cm^{-3}$ and different diffusion coefficient parameters ( $\chi=1$ and 0.01). A denser environment and lower diffusion coefficient can  enhance the flux of \gray line emission. 
The integrated fluxes of the narrow 4.44~MeV line emission, calculated for different combinations of $n$ and $\chi$, are presented in  Table \ref{tab-lflux}. The 4.44~MeV $\gamma$-ray line flux sensitivity of the proposed telescopes e-ASTROGAM \citep{e-astrogam2018} and AMEGO  \citep{amego2019} is $\sim 10^{-7}\, {\rm ph\,cm^{-2}\,s^{-1}}$  for the observation time $T_{\rm obs}=$1~yr.  The comparison with Table \ref{tab-lflux} shows that in very dense environments with $n\geq100\,\rm cm^{-3}$ and slow particle diffusion, the LECR sources can be revealed through $\gamma$-ray lines provided that the parameter $Q/d^2$ is not significantly smaller than $10^{38}\,\rm{erg\,s^{-1}\,kpc^{-2}}$.
In all figures above, the composition of the ambient medium is fixed to the solar abundance. However, in certain astronomical locations, the composition might be different, in particular, it might be enhanced by heavy elements in metal-rich environments, such as the Galactic center \citep{Benhabiles2013}, or in the young SNR Cas~A \citep{Summa2011casA}.  The higher abundance can dramatically enhance the $\gamma$-ray line fluxes and makes these sources prime targets for observations with next-generation MeV $\gamma$-ray detectors.  

\begin{figure*}
    \centering
    \includegraphics[width=0.45\textwidth]{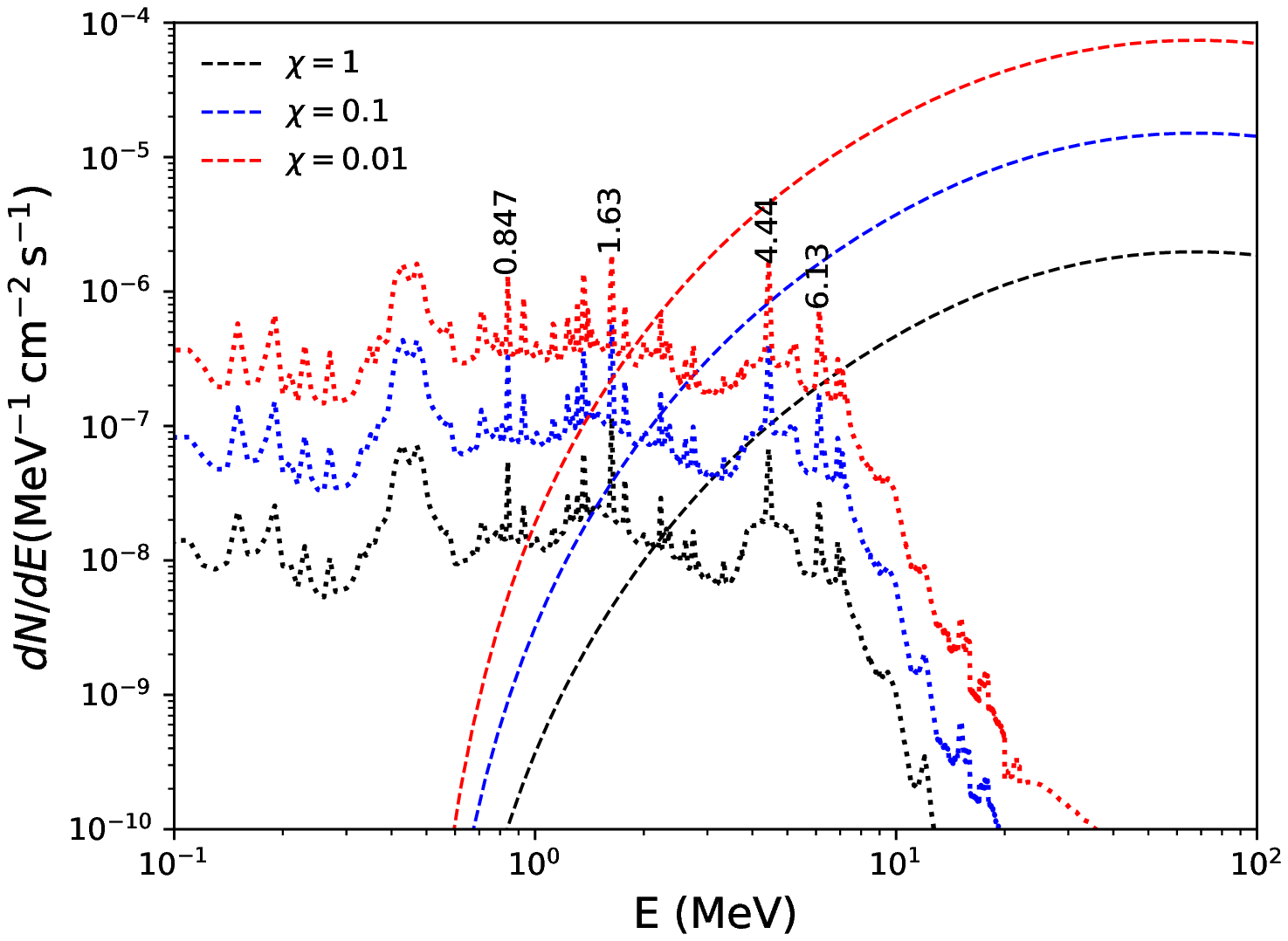}
    \includegraphics[width=0.45\textwidth]{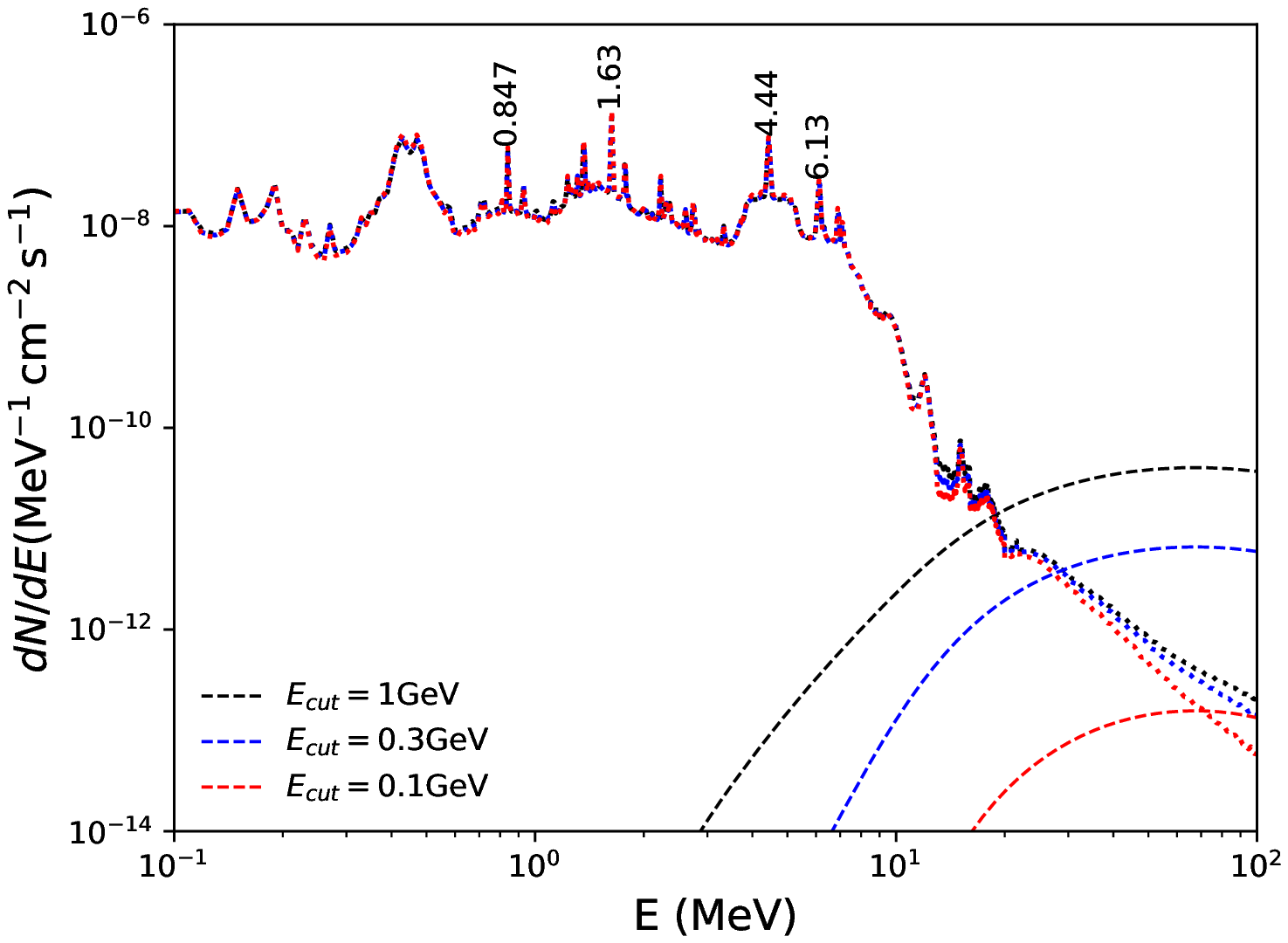}
\caption{Comparison of $\gamma$-ray fluxes resulting from de-excitation of nuclei (dotted lines) and  $\pi^0$-decay process (dashed lines) integrated within 25\,pc from the hypothetical CR source. 
In the left panel, we assume a proton spectrum $F_{\rm p}(E)\propto E^{-2}$ with various diffusion coefficient parameters, in which $\chi=1.0$ (black), 0.1 (blue), or 0.01 (red). In the right panel, we assume $\chi=1.0$ and proton spectra $F_{\rm p}(E)\propto E^{-2}{exp}(-E/E_{\rm cut})$ with various cutoff energies, in which $E_{\rm cut}=$ 1\,GeV (black), 300\,MeV (blue), or 100\,MeV (red).   
Meanwhile, $Q/d^2=10^{38} {\rm erg\,s^{-1}\,kpc^{-2}}$ and $n=1\,\rm cm^{-3}$ are assumed in both panels.
} 
\label{fig7}
\end{figure*}

So far, we did not include in calculations the $\gamma$-ray channel linked to  
$pp$ interactions with the production and decay of $\pi^0$ mesons. The relative contribution of this channel strongly depends on the proton spectrum, especially on the continuation of the proton spectrum beyond the kinematic threshold of $\pi$-meson production.  Setting $Q/d^2=1\times10^{38}\,\rm{erg\,s^{-1}\,kpc^{-2}}$ and $n=1\,\rm cm^{-3}$, in Fig.\ref{fig7} we show the differential spectra of $\gamma$-radiation consisting of the nuclear de-excitation lines and the $\pi^0$-decay $\gamma$-rays integrated within the region of radius 25\,pc around the source, assuming that the initial proton spectrum has a simple power-law distribution with index $s=2$ without cutoff (the left panel) or with an exponential cutoff at $E_{\rm cut}=$100\,MeV, 300\,MeV, and 1\,GeV (the right panel). The left panel of Fig.\ref{fig7} shows that  $\pi^0$-decay \grays exceed the luminosity of  nuclear lines. 
The reason is that the energy losses cause the spectrum of CRs below 100 MeV to become very hard, thus the production rate of nuclear lines is suppressed.  On the other hand, the emissivity  of $\pi^0$-decay $\gamma$-rays is dramatically suppressed when the cutoff in the spectrum of LECRs is close to the threshold of $\pi^0$-decay production around $280\,\rm MeV$, as demonstrated in the right panel of Fig.\ref{fig7}.

It is also interesting to compare our results with those in \citet{Benhabiles2013}, which were calculated for the inner Galaxy.  The results in \citet{Benhabiles2013} reveal a significantly sharper profile than our results. The differences come mainly from the twice solar abundance used in their calculation, and thus more contribution from the {\it \textup{direct}} processes. \citet{Benhabiles2013} moreover calculated a diffuse emission in a  much larger region, in which the variation in the CR spectrum on a small scale may have only a minor effect on the final results. The limited angular resolution of the current and planned MeV instruments means that such diffuse emission may be a better target than point sources.  On the other hand, our calculations treated the  spectral  and density variation of CRs carefully and were aimed at the nearby CR sources.   

\begin{table*}
\caption{Elemental composition }
\centering
\begin{tabular}{l c ccc}
\hline
Z & Element & Local CR\tablefootmark{a} &Solar \tablefootmark{b} \\
\hline
1& H&  1 & 1 \\
2& He&  $8.140\times10^{-2}$&$8.414\times10^{-2}$ \\
6& C &  $1.671\times10^{-3}$&$2.455\times10^{-4}$\\
7& N &  $2.444\times10^{-4}$&$7.244\times10^{-5}$\\
8& O &  $1.570\times10^{-3}$&$5.370\times10^{-4}$\\
10& Ne &  $1.507\times10^{-4}$&$1.122\times10^{-4}$\\
11& Na &  $1.784\times10^{-5}$&$1.950\times10^{-6}$\\
12& Mg&  $2.264\times10^{-4}$& $3.467\times10^{-5}$\\
13& Al& $3.302\times10^{-5}$& $2.884\times10^{-6}$\\
14& Si&  $1.898\times10^{-4}$& $3.388\times10^{-5}$\\
16& S& $2.087\times10^{-5}$& $1.445\times10^{-5}$\\
18& Ar&$4.554\times10^{-6}$& $3.162\times10^{-6}$\\
20& Ca&$1.195\times10^{-5}$&  $2.042\times10^{-6}$\\
26& Fe&$1.152\times10^{-4}$& $2.884\times10^{-5}$ \\
28& Ni& $6.452\times10^{-6}$&  $1.660\times10^{-6}$\\
\hline
\end{tabular}
\tablefoot{
\tablefoottext{a}{ The LECR abundance according to the \voy measurement  \citep[see][Table~3]{Cummings2016}.}
\tablefoottext{b}{ The recommended present-day solar abundances extracted from the Table~6 of \citet{Lodders2010}}
}
\label{tab-ab}
\end{table*}
\begin{table*}
\caption{Integrated 4.44 MeV line flux under different assumptions of $n$ and $\chi$}
\centering
\begin{tabular}{ c c| c}
\hline
 $n$ & $\chi$ &Flux  \\
  ($\rm cm^{-3}$)&& (${\rm photon\,cm^{-2}\,s^{-1}}$) \\
\hline
1&1.0&$5.49\times10^{-9}$\\
1&0.1&$3.25\times10^{-8}$ \\
1&0.01&$1.25\times10^{-7}$\\
100&1.0& $1.25\times10^{-7}$\\
100&0.1&$3.16\times10^{-7}$ \\
100&0.01&$5.96\times10^{-7}$\\
\hline
\end{tabular}
\tablefoot{
The injection rate of protons $Q/d^2=10^{38} {\rm erg\,s^{-1}\,kpc^{-2}}$ and spectral index $s=2.0$ are assumed. The FWHM width of 4.44 MeV narrow line $\Delta E$ is $\sim100$\,keV. Details is described in Sec.\ref{subsec:dis}.}
\label{tab-lflux}
\end{table*}

\section{Summary}
\label{sec:sum}

Together with magnetic fields and turbulent gas motions, Galactic CRs play a dominant role in the energy balance of the interstellar medium. Although subrelativistic (suprathermal) particles supply a substantial fraction of the CR pressure,  the real contribution of LECRs remains uncertain. Because of the slow propagation and severe energy losses, the local LECRs make negligible contributions to the fluxes beyond 100\,pc. For the same reason, LECRs are expected to be inhomogeneously distributed in the Galactic disk; subrelativistic protons and nuclei are expected to be concentrated around their acceleration sites. Because all potential CR source populations (SNRs, stellar clusters, individual stars, etc.) are linked in one way or another to the star-forming region, the effective confinement of LECRs in these regions is expected to produce feedback that stimulates the star formation through the ionization of the nearby molecular clouds. LECRs play a significant role also in the chemistry of the interstellar medium.  Thus an unbiased, observation based information about LECRs at the sites of their concentrations in the Milky Way is crucial for understanding the fundamental processes linked to the dynamics and chemistry of the interstellar medium, star formation, etc. The most direct channel of information is provided by nuclear de-excitation \gray lines resulting from the interactions of protons and nuclei of LECRs with the ambient gas. 

We explored the effect of the initial spectra shape, energy loss, and diffusion coefficient on the spatial and energy distributions of LECRs around the source that continuously injects accelerated protons and nuclei into the surrounding medium. LECRs, lose much energy during propagation, especially at energies below 100 MeV/nuc. At large distances, depending on the diffusion coefficient, their radial profiles become steeper than $1/r$ (see Fig.\ref{fig-pspec1}), which is the typical radial distribution of particles at higher energies at which the energy losses can be neglected. We calculated that characteristics of the nuclear $\gamma$-ray line emission initiated by interactions of LECRs with the ambient gas using the cross sections gathered from experimental data and theoretical modeling with the code TALYS-1.95. We found that for the standard diffusion coefficient characterizing the propagation of CRs in the interstellar medium, the \gray emission is mainly produced in regions around the source within $\sim25$\,pc. The slower diffusion with the parameter $\chi < 1$ 
makes the $\gamma$-ray source more compact and brighter. The expected $\gamma$-ray fluxes are unfortunately well below the sensitivity of current $\gamma$-ray detectors, however. With the arrival of the proposed $\gamma$-ray detectors that are dedicated to low energies, such as e-ASTROGAM \citep{e-astrogam2018} and AEMGO \citep{amego2019}, the detection of nuclear $\gamma$-rays, in particular, the lines at 0.847, 1.63, 4.44, and 6.13 MeV of nearby  (within a few kpc) accelerators of LECRs would become feasible provided that the LECR accelerators are surrounded by dense ($\geq 100\,\rm cm^{-3}$) gaseous regions in which LECRs propagate significantly more slowly than in the interstellar medium, $\chi \ll 1$.

\begin{acknowledgements}
Bing Liu  thanks Jurgen Kiener for providing very helpful information on the modification of TALYS structure files, and is supported by the Fundamental Research Funds for the Central Universities. 
Ruizhi Yang is supported by the NSFC under grants 11421303 and the national youth thousand talents program in China.
\end{acknowledgements}

\bibliographystyle{aa}

\end{document}